# Deterministic Printing of Single Quantum Dots

*Gregory G. Guymon[1], Hao A. Nguyen[2], David Sharp[3], Tommy Nguyen[3], Henry Lei[4], David S. Ginger[2], Kai-Mei C. Fu[3,5,6], Arka Majumdar[3,5], Brandi M. Cossairt[2], and J. Devin MacKenzie[1,4,7]\**

**Affiliations**

[1]Mechanical Engineering Department, University of Washington; Seattle, 98195, USA.

[2]Department of Chemistry, University of Washington; Seattle, 98195, USA.

[3]Department of Physics, University of Washington; Seattle, 98195, USA.

[4]Materials Science Department, University of Washington; Seattle, 98195, USA.

[5]Department of Electrical and Computer Engineering, University of Washington; Seattle, 98195, USA.

[6]Physical Sciences Division, Pacific Northwest National Laboratory, Richland, 99352, USA.

[7]Washington Clean Energy Testbeds, University of Washington; Seattle, 98105, USA.

*Corresponding author (Email: jdmacken@uw.edu)

**Funding**:

National Science Foundation (NSF) Center for Integration of Modern Optoelectronic Materials on Demand (IMOD), an NSF Science and Technology Center, under grant agreement DMR-2019444 (JDM, BMC, AM, KCF, DSG).

Molecular Analysis Facility, which is supported in part by funds from the Molecular Engineering & Sciences Institute, the Clean Energy Institute, and the National Science Foundation (NNCI-2025489 and NNCI-1542101)

Washington Research Foundation (AM, KCF, JDM)

Joint Center for Deployment and Research in Earth Abundant Materials (JDM)

Washington Clean Energy Testbeds, a facility operated by the University of Washington Clean Energy Institute,

Air Force Research Laboratory with SEMI-FlexTech (JDM, HL) under Agreement Number FA8650-20-2-5506 in support of Army Research Laboratory.

**Abstract**

The unique optical properties of quantum dots (QDs), size-tunable emission and high quantum yield, make them ideal candidates for applications in secure quantum communication, quantum computing, targeted single-cell and molecular tagging, and sensing. Scalable and deterministic heterointegration strategies for single QDs have, however, remained largely out of reach due to inherent material incompatibilities with conventional semiconductor manufacturing processes. To advance scalable photonic quantum device architectures, it is therefore crucial to adopt placement and heterointegration strategies that can address these challenges. Here, we present a electrohydrodynamic (EHD) printing model, single particle extraction electrodynamics (SPEED) printing, that exploits a novel regime of nanoscale dielectrophoretics to print and deterministically position single colloidal QDs. Using QDs solubilized in apolar solvents, this additive, near zero-waste nanomanufacturing process overcomes continuum fluid surface energetics and stochastic imprecision that limited previous colloidal deposition strategies, achieving selective extraction and deposition of individual QDs at sub-zeptoliter volumes. Photoluminescence and autocorrelation function ($g^{(2)}$) measurements confirm nanophotonic cavity-QD integration and single-photon emission from single printed QDs. By enabling deterministic placement of single quantum dots, this method provides a powerful, scalable, and sustainable platform for integrating complex photonic circuits and quantum light sources with nanoscale precision.

**Introduction**

The proliferation of electronic and photonic devices necessitates innovations in precise, scalable, and less wasteful fabrication of nanoscale components. Among techniques that can directly integrate device elements with diverse physical properties and compositions, additive manufacturing through inkjet printing has emerged as a versatile method for depositing picoliter liquid volumes, with applications ranging from graphic printing to advanced microelectronics[1,2] and photonics[3]. Printed electronics are particularly materials-efficient as compared to conventional subtractive processing of electronics[4]. Considering the emissions associated with processing alone, printing of electronics has been modeled to have more than two orders of magnitude lower environmental emissions footprint than conventional approaches[5]. Electrohydrodynamic (EHD) printing advances inkjet methods by employing electric fields, as opposed to mechanical or thermal stimuli, to manipulate fluid inks. EHD printing can readily achieve patterned feature resolutions well below 10 μm (**Figure 1a**)[6–14], enable attoliter-scale droplet deposition, and facilitate integration of functional materials onto a variety of substrates. Unlike conventional inkjet droplet formation mechanisms, EHD inkjet uses electric fields acting on ink components to overcome fluid surface tension barriers that limit the feature sizes limits of conventional approaches. This enables nanoscale control over deposit size and placement through a transfer mechanism that requires no moving parts, and is readily scalable into parallel printing modules for higher throughput manufacturing[15,16].

Commercialized drop-on-demand inkjet heads based on silicon micro-electromechanical systems (MEMS) that require complex integration of piezoelectric elements and tight control of nozzle geometry have been scaled to 2048 nozzles monolithically-integrated into single print heads[17]. Even for printing cycle rates as low as 1 Hz per ejected deposit, a similarly manufactured EHD print system, that requires only the application of electric fields directly to the printing ink could deposit thousands of QDs per second from a single head. Parallelized, multiple printhead systems with thousands of nozzles per head running with nozzle firing rates in the kHz range are now common in industrial piezoelectric inkjet printers. These systems are currently used in manufacturing of large area OLED displays with manufacturing throughputs of 60 seconds per square meter-scale sheet[18]. Research in high-resolution EHD printing has shown that kHz droplet printing rates can also be achieved with EHD at the micron and submicron scale[19]. The scaling of the number of nozzles per EHD print head is also progressing, as reflected in the recent announcement of 128 nozzle MEMS-fabricated EHD print heads operating in the kHz regime[20] and in the current installment of a 128 nozzle capable system at the University of Washington. It is, therefore, reasonable that industrial deployment of multi-head and multi-nozzle printing heads firing at kHz rates could be used in the manufacture of, for example, arrays of single QD light sources in an additive, low waste, low temperature, ambient pressure, digitally-controllable process.

Recently, by leveraging dielectrophoretic (DEP) forces, EHD printing has been shown to form and eject sub-attoliter liquid droplets carrying suspended nanoscale solids by polarizing the particles themselves in nonuniform electric fields[14,21]. This mechanism generates directional forces on the particles, driving solid-laden fluid along electric field gradients to allow the formation and acceleration of high surface energy small droplets towards a target substrate. Conventionally, the droplet diameters considered have been significantly larger than the suspended particles and the ink fluid has largely been treated as a continuum of liquid and solid constituents. In these cases, reduction of printhead nozzle diameters was leveraged to reduce droplet volumes and nanoparticle counts[9]. Here, we demonstrate a new regime of EHD printing, acting on sub-zeptoliter, highly-polarizable single particles, a method we refer to as single particle extraction electrohydrodynamics (SPEED) printing, to selectively extract and print individual quantum dots (QDs), a critical milestone for scalable integration of nanoscale materials. These semiconductor nanocrystals, known for their size-dependent properties, are desirable for next-generation technologies, including optical quantum devices[22,23], nano-transistors[24,25], sensors[26], light-emitting diodes (LEDs)[27,28], single-photon sources[29,30], and CMOS circuits[31]. As of recently, the incorporation of QDs into EHD-printed LEDs[32,33], photon-detectors[34,35], and field-effect transistors[36] demonstrates that the technique is increasingly relevant across a growing spectrum of applications.

For this study, we selected ligand-dispersed CdSe QDs with colossal CdS shells (80 monolayers) due to their relatively large size (~70 nm in diameter) and high dielectric constant that result in large static electric dipole moments, good dispersion, and optoelectronic stability[37,38]. In the new EHD printing regime explored here, we have defined electric field and waveform conditions, modeled the electric field environment enabling single QD printing, and determined a QD size range for a given applied electric field to enable selective deposition of single nanoparticles. Furthermore, we have employed this EHD printing mechanism, along with spatial positioning and optical alignment hardware, to pattern single QDs in arrays and achieved deterministic placement of single QDs on nanophotonic cavities. We have also characterized the optical properties of the printed QDs and demonstrated, for the first time, single-photon emission from deterministically printed QDs in arrays and coupled to an optical cavity. We propose SPEED printing as a versatile method for selectively depositing single nanoparticles, offering a transformative pathway for fabricating advanced classical and quantum electronic and photonic devices with nanometer-scale resolution.

**Results**

*DEP principle and colloidal ink design*

**Figure 1b** schematically illustrates our EHD printing setup for positioning single QD particles. Printing is initiated by applying a voltage between the electrode-integrated glass capillary printhead and the substrate. This creates an electric field that guides the QDs to the substrate. The top inset for Figure 1b shows a fluorescence microscopy (FLM) image of a capillary printhead loaded with dilute colossal QD suspension showing the presence of isolated emitters that include dispersed single QDs. By varying the printhead-to-substrate standoff height and applied voltage, the electric field intensity can be maximized at the tip interface. Colloidal particles in an "ink" composed of oleic acid capped CdSe/CdS core/shell QDs suspended in an apolar solution mixture are then driven to the tip interface by the electric field. This particle mobility is achieved by utilizing a medium with a lower relative permittivity than that of the dispersed particles, which gives rise to a polarization disparity between the two that forces the particles towards regions of high electric field intensity. The QDs used in this work were synthesized via a volumetrically tunable strategy that enables the QD core to maintain its quantum-confined optoelectronic characteristics, while increasing the overall size and dipole moment of the particle in an applied field through the shell design (**Figure 1c**). The tunable polarization is crucial for generating the DEP forces needed to overcome the net interfacial energies of a single particle, especially when factors like electric field intensity and material properties have finite limits, as discussed in the following sections. The colossal core/shell synthesis yields slightly asymmetric, hexagonal, faceted particles that we approximate as spherical for later numerical and analytical estimations, due to their relatively isotropic morphology[39]. Our simulation shows that the particle asymmetry is not likely to significantly change the printing behavior here (**Figure S1**).

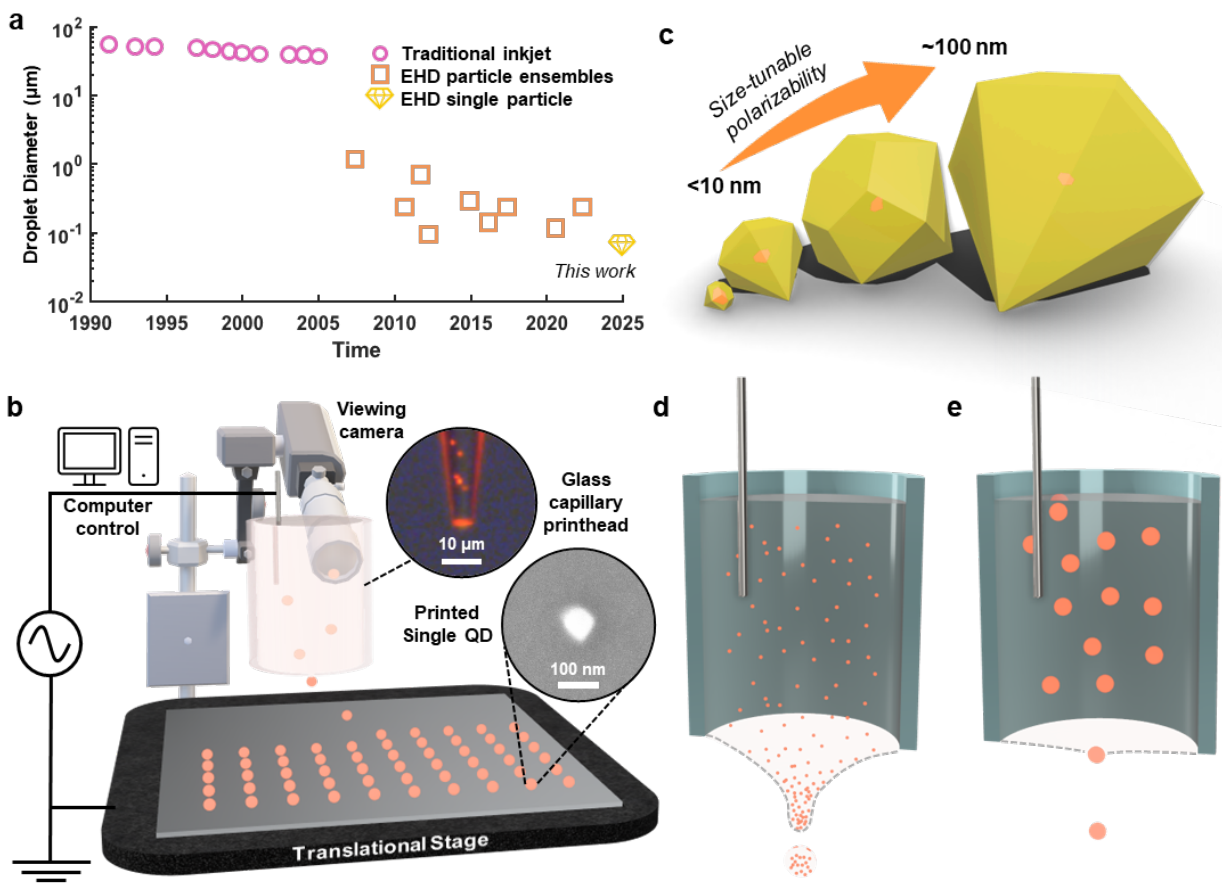

**Figure 1: Overview of experiment. a**, Reported droplet sizes using conventional mechanical inkjet and EHD printing over the past 35 years, along with the results of this work. **b**, Schematic diagram of the EHD printing setup used here. Top inset: FLM of EHD printhead. Bottom inset: Scanning electron microscopy (SEM) of single EHD-printed QD. **c**, Graphic representation of CdSe QD cores (orange) with tunable range of hexagonal diamond CdS shell (yellow) diameters. **d** and **e**, Bisected illustrations of EHD printheads dielectrophoretically overcoming (d) surface tension to print droplets of particle ensembles and (e) interfacial forces to print singular particles.

*Interfacial energy analysis*

In order to determine the optimal EHD printing and material design parameters, we first approach the process of separating an individual nanoparticle from a liquid-solid interface by first principles. Each interface in a three-phase system has an associated energy that quantifies the intermolecular interactions dictating the work to expand said interface. These energies can be described as the surface free energy ($\sigma_{SG}$), surface tension ($\sigma_{LG}$), and interfacial tension ($\sigma_{LS}$), which correspond to the solid-gas, liquid-gas, and liquid-solid interfaces, respectively. By considering the contact angle ($\theta_{CA}$) at the liquid-gas interface, the values can be related to one another via Young's equation:

$$\sigma_{LS} = \sigma_{SG} - \sigma_{LG} \cos(\theta_{CA}) \qquad (1)$$

Conventional EHD printing strategies focus on overcoming the energy of surface tension (**Figure 1d**) to generate higher area, reversed curvature liquid-gas interfaces in the form of convex Taylor cones and droplets. Alternatively, to separate a particle from a liquid-solid interface, our focus is on overcoming interfacial tension (**Figure 1e**). This energy determines the work ($W$) needed to deform the liquid interface around a particle, where $\Delta A$ represents the change in interfacial area as the particle passes through the meniscus interface:

$$W = \sigma \Delta A \qquad (2)$$

The total energy required can be determined by considering the net positive and negative changes in the interfacial area for each interface. For a spherical model, the force ($F_{net}$) required to overcome the net interfacial forces can then be derived with respect to the translational z-axis, i.e. the printing direction towards the target substrate, where $a$ is particle radius:

$$\frac{dW}{dz} = F_{net} = (\sigma_{SG} - \sigma_{LS} - \sigma_{LG})2\pi a \qquad (3)$$

We measured the components of interfacial tension through optical tensiometry to determine the $F_{net}$ as a function of particle perimeter (see methods). First, we performed a pendant drop test (**Figure S2**) to determine the surface tension of the apolar solution, which agrees with previously measured values of the solution's constituents[40]. For the free surface energy of the QDs, we performed a sessile drop test on prepared QD thin films (**Figure 2a**), which also agrees with former analysis of the QD ligand chemistry[41]. We then solved for the interfacial energy in equation (1), resulting in a value of $6.9 \pm 1.0$ mN/m, which allows us to estimate the $F_{net}$ a single particle needs to overcome to print.

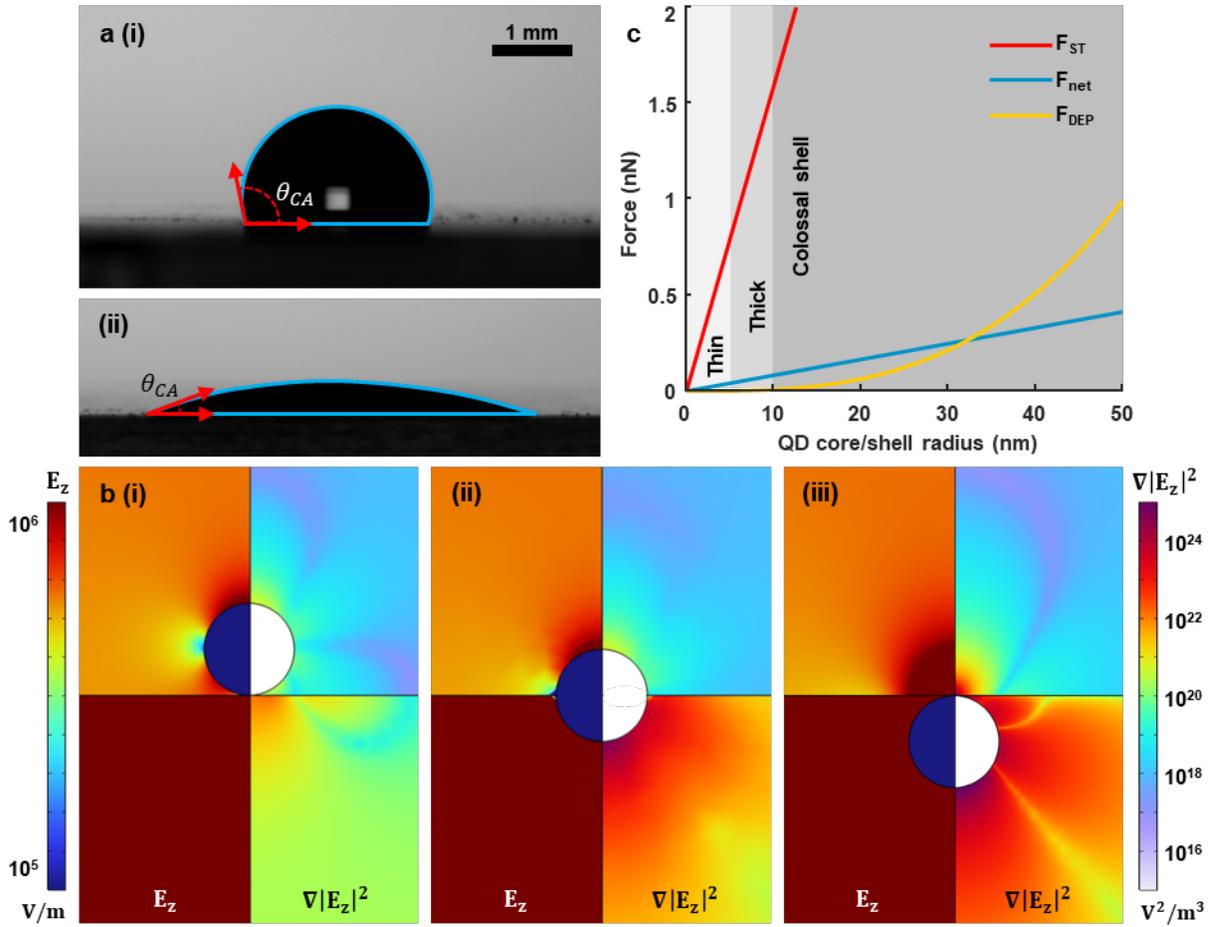

**Figure 2: Characterization of ink and energy modeling. a**, Shadow images of sessile drop shape analysis of (i) water and (ii) octane:hexadecane mixture on CdSe/CdS QD thin film. Sessile droplet outlines (blue) indicate shape profile used for contact angle and surface free energy measurements. **b**, Electrical COMSOL simulation of spherical particle at the (i) interior solution interface, (ii) halfway through solution interface, and the (iii) exterior of the solution interface. The left side of each plot shows the z-component of the electric field ($E_z$) and the right side shows the z-component of the gradient of the field magnitude squared ($\nabla |E_z|^2$), (the z-axis is normal to the interface). The simulation is symmetrical around the center axis. **c**, Plot of surface tension force (red), net interfacial forces (blue) and DEP forces (yellow) acting on single particles of various radii. The shaded background indicates the size regime of the CdS shells.

*Single nanoparticle printing mechanism*

The DEP force ($F_{DEP}$) of a spherical particle under an applied non-uniform AC field can be solved analytically[42]:

$$F_{DEP} = \pi \varepsilon_m a^3 \Re[k(\omega)] \nabla |E|^2 \qquad (4)$$

where $\varepsilon_m$ is the real permittivity of the medium and $\nabla|E|^2$ is the gradient of the field magnitude squared for the particle. With the complex permittivities of the particle ($\tilde{\varepsilon}_p$) and medium ($\tilde{\varepsilon}_m$), the direction and magnitude of the force is described by $k(\omega)$, the frequency-dependent Clausius-Mossotti factor:

$$k(\omega) = \frac{\tilde{\varepsilon}_p - \tilde{\varepsilon}_m}{\tilde{\varepsilon}_p + 2\tilde{\varepsilon}_m} \qquad (5)$$

Simply put, when the particle is more polarizable than its respective medium, like in the approach we propose here, the particle moves along the electric field gradients towards regions of high field strength, often referred to as positive DEP force, and displaces the less polarizable solvent medium. For QDs with separate core/shell materials, the real component of $\tilde{\varepsilon}_p$ can be calculated as the volumetric ratio between the shell and core's real permittivities[43–45]. As the shell radius increases past the thin (<5 nm) to the thick (5-10 nm) and into the colossal (10-50 nm) regimes[37], the contribution of the core to $\Re[k(\omega)]$ becomes negligible, asymptotically approaching $\Re[k(\omega)] \approx 0.52$ (**Figure S3**), assuming a fixed medium permittivity[46,47].

We estimate the $\nabla|E|^2$ at the QD particle at the interface of the printhead's fluid meniscus in our EHD printer to be on the order of $\geq 10^{23}$ V²/m³ using a geometric electrical model by numerically solving Maxwell's equations (COMSOL, **Figure 2b**). Using the measured dimensions of the EHD printhead (**Figure S4**), we simulated the environment around the particle using the maximum voltage that our setup can operate to assess its feasibility (1 kV bias and 1kV amplitude).

Here, we assume the conductivity of the particle to be similar to CdS thin films, $10^{-7} \sim 10^{-9}$ $\Omega^{-1}$-cm$^{-1}$, as the CdS constitutes the majority of the material volume at large shell radii[48–50] and the shell diameter is significantly larger than Bohr radius for CdS[51]. The plotted z-component of the electric field ($E_z$) shows a high contrast field intensity difference between the solution and air, which draws the particle towards the higher field intensity. Additionally, the z-component of the gradient of the field magnitude squared plot shows significant field enhancement at the particle, increasing as it passes through the solution interface. Using these estimates, **Figure 2c** compares $F_{DEP}$ to $F_{net}$ and pure surface tension force ($F_{ST}$). Here, we see that $F_{DEP}$ grows exponentially with increasing particle radius, overcoming net interfacial forces as the particle enters the colossal regime. However, the DEP forces at this size are significantly less than what would be expected to form a liquid droplet of equivalent size from solution if we treated the ink solution as a continuum fluid as in conventional models for EHD inkjet. By comparing the competing forces, we estimate that a minimum QD radius ~32 nm would be sufficient to overcome the interfacial forces to separate a single QD with our measured interfacial energy from the liquid. To be able to extract and print a single QD and maximize the colloidal stability of these QD solutions, we selected a target particle radius of ~35 nm for further experimentation.

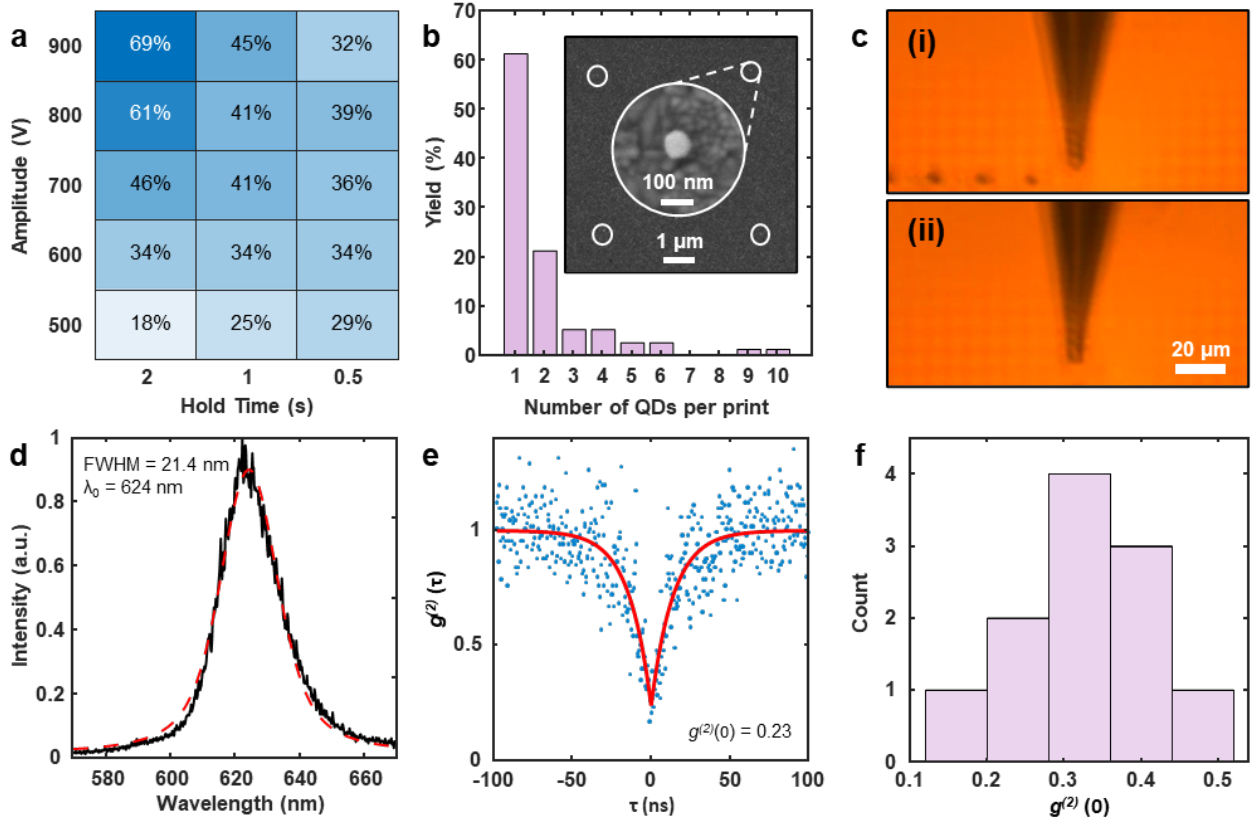

**Figure 3: SPEED printing of single QDs. a**, Effect of drive amplitude on FLM detected print yield as a heatmap of number of prints detected out of a 10 x 10 array of print attempt sites, per parameter configuration. **b**, Histogram plot of QDs counted per print site for a 10 x 10 array (900 V and 2 s parameter configuration) of attempt sites, as determined by SEM. Inset: SEM image of 2 x 2 array of SPEED-printed single QDs on ITO-glass. White circles indicate single QD positions. Central inset: SEM image of single QD. **c**, Observation camera image stills of EHD printhead printing (i) thin-shelled QDs and (ii) colossal-shelled QDs. **d**, Plot of single QD emission spectrum (black) and Lorentzian fit (red). **e**, Second-order time correlation plot (blue) and fit (red) of SPEED-printed QD showing single-photon emission, measured at room temperature, with 10 μW excitation (d and e). **f**, Histogram plot of $g^{(2)}(0)$ measurements on SPEED-printed single QDs (n = 11).

*SPEED printing single QDs*

To test this estimate, we synthesized CdSe/CdS core/shell QDs with a calculated core radius of 1.8 nm and a designed total average radius of $36 \pm 3.1$ nm from 80 CdS shell monolayers (see methods) and dispersed the particles in octane:hexadecane with phosphonic and carboxylic ligands to stabilize solubility. **Figure 3a** shows the results of a printing trial matrix with this dispersion on Indium Tin Oxide (ITO)-coated glass across a variety of print electrode voltage amplitudes (500-900 V) and hold times (0.5-2 s), with a fixed bias (1 kV) and frequency (1 kHz). Here, the hold time refers to the period that the amplitude voltage is applied over a target position. Afterwards, using machine vision to count the number of detectable prints (**Figure S5**), the heatmap shows a gradient of successful QD prints, as determined by FLM. As the hold time and amplitude voltage increase, the number of prints made relative to attempts increases. This trend is explained by the effects of competing forces on the particle's overall force magnitude normal to the meniscus as the QD reaches the interface. The viscous Stokes drag force ($F_\eta$) on a particle is equal to the $F_{DEP}$ at terminal velocity in the bulk solution until reaching the interface[43]. As the particle approaches the meniscus interface, when $F_{DEP} \geq F_{net}$, the particle decelerates, decreasing $F_\eta$ until $F_{DEP} = F_{net} + F_\eta$. At this point, the particle continues to move through the interface according to Newton's second law. Thus, increasing the hold time and/or strength of electric field translates to a longer time for the ejection to occur and faster ejections, respectively. It should be noted that during this printing process, we could not discern visible meniscus perturbations (Figure 3c(ii)) through high-magnification viewing cameras mounted on the print system (**Figure S6**), indicating negligible effects of the applied electric fields on the apolar solvent. This, in combination with the approximately 5 μm printing nozzle orifice diameter, which is much larger than the printed QDs, indicates that this nanoscale printing is decoupled from the nozzle size.

Examining the highest print yield array (900 V and 2 s) with SEM (**Figure S7** and **Extended Data 1**), **Figure 3b** summarizes the number of QDs detected per print via SEM, resulting in a single particle yield slightly over 60%. It's worth noting that over a quarter of the print sites with 2+ QDs showed intimate contact between QDs, suggesting that they were likely already aggregated before being printed. The remaining multiples were observed as sets of isolated single QDs, within a few hundred nm of one another at each print site. SEM also revealed single particles for positions with no emission (n = 7), indicating that Figure 3a underestimates the total printing yields of prints made. We attribute this lack of emission to either non-radiative recombination from excessive defects in the particles, photobleaching by excitation[52], or QD-excluded CdS aggregates. The inset of Figure 3b shows a 2 x 2 subarray of single QDs from the larger array. The SEMs show the expected faceted structure consistent with the known morphology of these colossal QDs[37]. We note the lack of any observable drying rings or residue in the vicinity of the QDs. Such features are typically observed when a liquid droplet wets a substrate[53]. The lack of meniscus deformation when compared to the printing behavior of conventional thin-shelled QDs seen in **Figure 3c**, in combination with (i) the high surface tension barrier for liquid droplet formation (Figure 2c, red curve), (ii) the calculated dominance of the dielectric forces for QD with a bulk dielectric constant of ~9 over the QD interfacial forces at the meniscus (Figure 2c, yellow and blue curves), and (iii) negligible interaction of the printing electric fields with the solvents themselves, with a dielectric constant of ~2, lays out a new model for EHD printing at the nanoscale. This mechanism is formalized as SPEED printing, describing the single-particle extraction of nanocrystals from solution without deforming a bulk liquid meniscus. Here, we propose the extraction and directed printing of dry or nearly dry core/shell QDs from apolar solutions with appropriate surface energies, dielectric properties, and electric field environments. COMSOL simulations at the experimented conditions (**Figure S8**), slightly below the printer's maximum voltages (Figure 3b), showed a negligible reduction in field intensity around the particle, showing that single particle printing can be achieved at lower voltages. Experiments with thin-shelled[21] and smaller colossal-shelled (**Figure S9**) CdSe/CdS QDs, failed to achieve individual particle printing, again consistent with the proposed model.

We also printed single QDs on silicon nitride-coated Si wafers to improve the signal-to-noise for spectral analysis. **Figure 3d** shows the room temperature photoluminescence (PL) of a single printed QD with an emission peak at 624 nm, and a full-width half-maximum of ~21 nm, consistent with the same QDs deposited by drop-casting approaches[37] and measured in solution (**Figure S10**). Second-order correlation ($g^{(2)}$) measurements confirm single-photon emission from printed QDs for the first time. Figure 3d shows an example of a QD maintaining its single-photon emissivity ($g^{(2)}(0) < 0.5$) after printing, a critical metric for utilizing these materials as single photon sources in quantum device technologies. Sampling an array of printed single QDs (**Figure 3f**) shows single-photon emission, with a mean $g^{(2)}(0) = 0.33 \pm 0.1$.

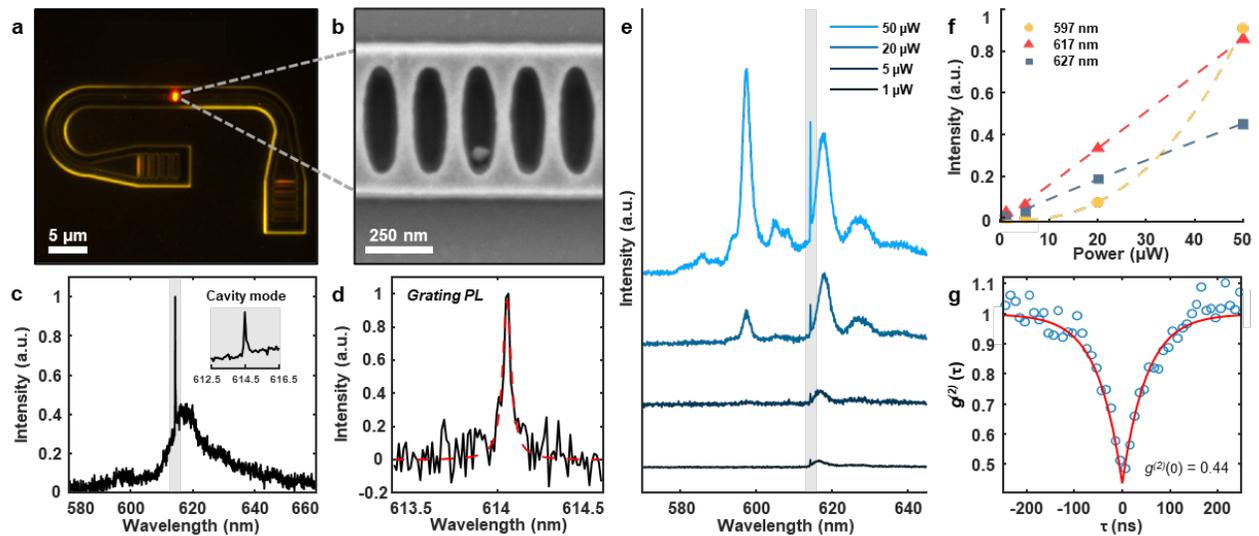

**Figure 4: Single QD nanophotonic cavity heterointegration. a**, Fluorescent microscopy of horseshoe-shaped cavity with single SPEED-printed QD. **b**, SEM image of the cavity region containing single QD. **c**, Spectrum plot of cavity integrated QD, at 70 K, excited and measured from the top. Inset: Magnified spectrum (shaded) around the cavity mode. **d**, Spectrum of the pure QD-coupled cavity mode PL to the grating (black) and Lorentzian fit (red) when exciting from the top, measured at room temperature with 10 μW excitation (C and D). **e**, Power-dependent spectra of QD-coupled cavity at 8 K with the cavity mode region shaded (614 nm), excited and measured from the top. **f**, Power dependence of 597 nm (yellow), 617 nm (red), and 627 nm (blue) peak intensities at 8 K, with power-law fits (dashed lines). **g**, Second-order time correlation plot (blue) and fit (red) of cavity-coupled QD, excited (10 μW) and measured from the top at room temperature.

*Nanophotonic cavity Heterointegration*

To demonstrate the utility of high-resolution single QD printing, we designed and fabricated a horseshoe-shaped nanophotonic cavity with a resonant frequency of ~614.5 nm and a monolithic waveguide structure that directs emission to rectangular gratings at each end. Applying the strategy described above, we printed a single QD in the center of the cavity. This was further enabled by the precision of the step motor driven stage, enabling step precision down to 100 nm (±5%) in lateral directions, sufficient for alignment to the ~500 nm critical dimension width of the nanophotonic device target. **Figure 4a** shows a FLM image of the emission from the single QD coupling to the waveguide and scattering from the gratings. SEM inspection of the print site (**Figure 4b**) confirms the presence of the single QD by identifying the characteristic hexagonal diamond-shaped colossal QD in one of the cavities' central elliptical wells.

The PL spectrum from the cavity-integrated QD in **Figure 4c** reveals the coupling between the cavity's sharp resonant mode and a broad QD emission at 70 K, a condition chosen for its high emission intensity (**Figure S11**). Examining the narrow spectrum around the cavity mode (Figure 4c, inset) is consistent with a modest Fano interference, indicated by the characteristic Fano line shape[54], which is more apparent at room temperature (**Figure S12**). Measuring the room temperature photoluminescence from the cavity grating, at either end of the horseshoe, isolates emission coupled to the cavity mode. **Figure 4d** shows this sharp Lorentzian mode, with a quality factor of 12,500, measured from the Lorentzian fit linewidth over the center wavelength. Similar to our printed arrays, photobleaching also occurred in other single QD-integrated cavities, resulting in only the emission from the SiN background and cavity mode (**Figure S13**).

We measured the QD-cavity system at 8 K across multiple excitation powers (1-50 µW) to examine the broadened tail emission. **Figure 4e** shows the separation of the broader emission into an excitonic peak (617 nm) and its phonon sideband (627 nm). At higher power (>5 µW), we identify the emergence of a secondary high-energy peak (597 nm), which we attribute to a multi-excitonic transition, accompanied by its own phonon sideband (605 nm). The observed single excitonic and phonon sideband emission is consistent with non-cavity QDs (**Figure S14**) and the literature[55], at low temperature. As shown in **Figure 4f**, the exciton and phonon sideband scale nearly linearly with excitation power. The high-energy peak is highly non-linear, with an exponent of 2.5, suggesting a multi-excitonic origin. Lastly, a $g^{(2)}(\tau)$ measurement of the QD-cavity at room temperature (Figure 4f) shows the coupled system maintaining its single-photon emission.

**Conclusion**

Colloidal QDs have been discussed for quantum optoelectronic applications since their inception yet have largely been outside of practical considerations due to the challenges associated with their integration into devices. Here, we have demonstrated the first additive nanomanufacturing strategy for deterministically positioning single QDs that exhibit single-photon emission. Not only does this approach allow high throughput printing of large arrays of QDs at room temperature, without resorting to vacuum processing or lithography, but it also allows for the positioned deposition of QDs into prefabricated nanophotonic elements such as high-Q cavities. We propose that this printing is enabled by a new electrohydrodynamic printing mechanism, single particle extraction electrohydrodynamics (SPEED) printing, defined by dielectrophoretic expulsion of essentially dry, highly polarizable single particles from a relatively non-interacting solvent medium. This mechanism requires particles with sufficient dielectric contrast from the solvent, which can be realized with typical semiconductor nanocrystals at sizes that are now synthetically achievable. Colossal-shelled, (~70 nm) CdSe/CdS particles with oleate surface ligands, and a bulk particle to solvent dielectric constant ratio, of 9:2 were used to demonstrate single particle printing here. We anticipate, however, that the method should be applicable to other colloidal nanoparticles, from halide perovskites, to oxides and nanodiamonds that can be synthesized into single or multi-component particles with a sufficiently high overall polarizability and particle to solvent dielectric contrast, size, appropriate surface energy and stability in apolar solvents such that the dielectrophoretic forces can overcome surface energetics as described in the Results discussion here. This SPEED printing process opens the door to a scalable, contactless, zero-wasted additive nanomanufacturing process that can be conducted at ambient conditions to enable the positioning of single discrete colloidal QDs. Even under early-stage, non-optimized conditions and chemistries, single-particle deterministic printing yields reached over 60%, validating the underlying mechanism and enabling pathway optimization moving forward. These results represent significant progress in realizing scalable photonic qubit platforms and the future development of quantum optoelectronics. We also believe this represents a step forward for the additive manufacturing of materials, where low-waste printing processes can match and even exceed the capabilities of conventional subtractive and complex semiconductor fabrication processes.

**Materials and Methods**

Chemicals and Materials

All chemicals listed below were used without further purification, unless stated otherwise. Oleic acid (≥90%), anhydrous methyl acetate (99.5%), selenium powder (99.999%, 100 mesh), cadmium oxide powder (CdO, 99.9%), trioctylphosphine oxide (TOPO, 99%), trioctylphosphine (TOP, 90%), 1-octanethiol (99%), 1-octadecene (ODE, 90%), n-octane (98%), hexadecane (99%), and anhydrous hexane (95%) were purchased from MilliporeSigma, and n-octadecylphosphonic acid (ODPA, 99%) was purchased from PCI Synthesis. Cadmium oleate solution (0.2 M in ODE) was made by heating CdO (2.57 g), oleic acid (63 mL), and ODE (27 mL) at 160 °C for 30 min, followed by evacuating the solution at 110 °C for 1 h.

Methods

*CdSe QD Synthesis*

The synthesis of colossal CdSe/80CdS QDs (CdSe with 80 monolayers (ML) of CdS) was conducted according to methods reported previously[37]. The wurtzite CdSe core synthesis started with adding TOPO (3 g), CdO (0.06 g), ODPA (0.28 g), and a magnetic stir bar to a 15 mL three-neck round-bottom flask secured with a septum, a thermowell equipped with a thermometer, and a condenser. The flask was put under vacuum for 30 min at 100 °C, then was refilled with N2. The temperature was then set to 375 °C. A dropwise addition of TOP (1.5 mL) was done while the temperature was ramping up from 320 to 370 °C. When the temperature was at 375 °C, a pre-prepared solution of Se powder (0.058 g) dissolved in TOP (0.4 mL) was quickly injected. After 45 s, the flask was quickly cooled to room temperature with forced air to yield CdSe QDs. The QDs were purified in a glovebox by centrifugation with methyl acetate/hexane and were stored in hexane.

*Shelling of CdSe with CdS*

ODE (6 mL) was added into a 25 mL three-neck round-bottom flask secured with a septum, a thermowell equipped with a thermometer, and a condenser, which was put under vacuum at 110 °C for 60 min, then refilled with N2 gas. An amount of CdSe core in hexane (typically 5–100 nmol) was added to the flask. Hexane was completely removed by evacuating the flask 110 °C for 30. The flask was then heated to 310 °C. When the temperature was at 260 °C, 0.2 M Cd(oleate)$_2$ and 0.2 M 1-octanethiol solutions were slowly injected into the reaction flask using a dual-syringe pump with the rate of 3 mL/h. The volumes of the shell precursors were calculated to achieve 30 MLs of CdS shells on QD cores with a diameter of ~3.5 nm. After the injection finished, the reaction was annealed at 280 °C for 2 h, then cooled down to 60 °C. Without purification, a portion of the yellow CdSe30CdS solution was transferred to a different 3-neck round-bottom flask for another shelling step that targeted additional 20 CdS MLs to achieve CdSe50CdS using the same shelling procedure. This step was repeated one more time to achieve CdSe80CdS. The final product was purified in a glovebox with hexane and then stored in hexane. The colossal QDs used in this study are the same as those described in a previous report[37], which reports the synthesis and full morphological/structural characterization; in that work, an ~18% blinking fraction and an average photobleaching lifetime of ~570 s were measured, consistent with the occasional non-emissive or transiently emissive particles observed after printing.

*QD Ink Preparation*

The QD ink was prepared by evaporating a hexane-based QD solution under vacuum at 100 ºC for 30 min, yielding ~3 mg of bright yellow QD powder. The resulting powder was subsequently dispersed in 10 mL of a 1:1 octane:hexadecane mixture through stirring at 80 °C for 1 h, followed by sonication 1 h.

*Pendant Drop Tensiometry*

The video image was calibrated to the outer diameter of the dispensing syringe of a Krüss drop shape analyzer to establish the scale and obtain accurate drop dimensions. The shape of the drop was then analyzed through grayscale shadow image processing. A shape parameter was adjusted iteratively by a Young-Laplace numerical model until the computed drop shape matched the observed one. Finally, the surface tension was calculated, based on the solution density and the optimized shape parameter, to be $25.3 \pm 0.3$ mN/m.

*Sessile Drop Tensiometry*

A substrate for the sessile drop tensiometry was prepared via spin coating CdSe/CdS on a glass substrate. Using a Krüss drop shape analyzer, a drop was deposited onto the solid sample, and its image was captured using a camera, then processed through drop shape analysis software. The drop's contour was first identified using grayscale shadow image analysis. Next, a geometric model was applied to fit the identified contour. The contact angle was determined as the angle between the fitted drop shape and the surface of the sample. These were determined as $92.3° \pm 2.0°$ and $18.5° \pm 2.9°$ for water and 1:1 octane:hexadecane, respectively. After the drop shape and contact angle were measured, the surface free energy of the solid sample was calculated, using a combination of contact angles from multiple probe liquids (water and 1:1 octane:hexadecane) with known surface tensions, to be $30.8 \pm 1.0$ mN/m.

*Finite Element Analysis*

Simulation was performed with the electric current (EC) package within the AC/DC Module using COMSOL Multiphysics. Permittivity and conductivity of the materials were used to perform the simulation. Physical dimensions of the components were modeled after micrographs (printing nozzle), SEM characterization (quantum dot) and measured substrate thickness. A source voltage of 2000 V was used as a stimulus to the electrode voltage, and the printing bed was used as the reference ground for the applied voltage, both analogous to the actual printing setup. A 2D axial rotationally symmetric simulation was performed for the analysis. A flat meniscus was assumed for the liquid-gas interface.

*Electrohydrodynamic Inkjet Printing*

Electrohydrodynamic inkjet printing was conducted by loading the prepared CdSe/CdS ink into pulled borosilicate glass capillary pipettes with a tip internal nozzle diameter of ~5 µm into a SIJ-050 Super Inkjet Printer. These printheads were manufactured with a tungsten electrode filament integrated inside this pipette. After mounting the printhead into the printer and placing the substrate on a grounded stage vacuum chuck, printing could be initiated. For the nanophotonic cavity, an alignment process was used to map the designed printing pattern to the orientation of the target substrate. Upon completion, printing was initiated with the following parameters: bias: 1000 V, amplitude: 900-1000 V, frequency: 1 kHz, hold time 1-2 seconds.

*Photoluminescence Characterization of Bare Substrate QD*

Single-particle characterization was performed in air using a custom-built scanning PL set-up. A 532 nm continuous-wave laser (Laserglow Technologies 532 nm DPSS Laser) was focused on the particles with a 100× (NA 0.95) dry objective lens with a typical excitation power of 10 µW. The laser spot position was controlled via a steering mirror (Newport FSM-300–01). The collected

PL was filtered through both a 550 nm long-pass filter and a bandpass filtered centered on 618 nm with 50 nm bandwidth. The collected PL was fiber-coupled either to a spectrometer (Princeton Instruments Isoplane 100) to collect the single-particle spectrum or through a 50:50 fiber splitter to a pair of avalanche photodiodes (MPD-PDM) to collect the $g^{(2)}$ spectrum. For the $g^{(2)}$ spectra, the time correlation was performed via a time tagger (PicoQuant TimeHarp 260).

*Nanophotonic Cavity Fabrication*

The SiN nanobeam cavities were designed for operation on a $SiO_2$ substrate with no top cladding in 220 nm thick SiN. The cavity was formed by punching a one-dimensional array of elliptical holes in a 550 nm wide SiN waveguide. The dimensions of the elliptical holes were fixed to a major diameter of 411 nm and a minor diameter of 98 nm. Each half of the cavity consisted of 10 elliptical holes that quadratically tapered from a period of 184 nm to 190 nm. An additional 20 elliptical holes with a period of 193 nm were placed on either end to form the cavity mirrors. According to FDTD simulations, this resulted in a cavity mode at 618 nm with quality factor ~4 x $10^5$ and mode volume ~ 2 $(\lambda/n)^3$. To fabricate the nanobeam cavities, 100 kV ebeam-lithography was used to transfer the pattern to a positive-tone ebeam resist. The pattern was then etched into the SiN thin film using a fluorine-based plasma etch.

*Photoluminescence Characterization of Cavity-Integrated QD*

Measurements were performed in a confocal setup operating in ambient conditions. Green continuous wave 532nm laser light (LaserQuantum opus532) was passed through a 532 dichroic mirror (Chroma ZT532RDC) focused onto the center of the cavity using a 40x/0.6 NA Olympus objective, with a 1 μm spot size. A 580LP was used to filter out excitation light and PL went to a Princeton Instruments 300 grooves/mm grating spectrometer (SpectraPro HRS-750) coupled to a Pixis CCD (PIXIS: 100BR_eXcelon). Confocal scans were performed using a SPCM (Excelitas

SPCM-AQ4C) coupled to a NIDAQ (BNC-2110). g2 was measured by going through a 40:60 fiber beam splitter and a timetagger (Swabian Instruments Timetagger Ultra). All low temperature measurements were done with a Montana cryostat (cryostation S-series) operating at 8K. Peaks in Figure 4e were integrated and fit to a power law $I = aP^b$ (Figure 4f), with the emission intensities scaling with excitation power (P) : excitonic ($P^1$), multi-excitonic ($P^2$ or higher), and phonon-assisted ($P^1$ or higher).

**Data and materials availability:** All data are available in the main text or the supporting information.

**Acknowledgements:** We thank Ethan Schwartz for ITO/glass substrate preparation.

**Author contributions:**

Conceptualization: GGG, HL, JDM

Methodology: GGG, HAN, HL, JDM

Investigation: GGG, HAN, DS, TN, HL

Visualization: GGG, HAN, TN

Funding acquisition: JDM, BMC, AM, KCF, DSG

Project administration: JDM, BMC, AM, KCF

Supervision: JDM, BMC, AM, KCF

Writing – original draft: GGG, HL, JDM

Writing – review & editing: GGG, HAN, DS, TN, HL, DSG, KCF, AM, BMC, JDM

**Competing interests:** Authors declare that they have no competing interests.

**Extended Data 1 - Row 1**

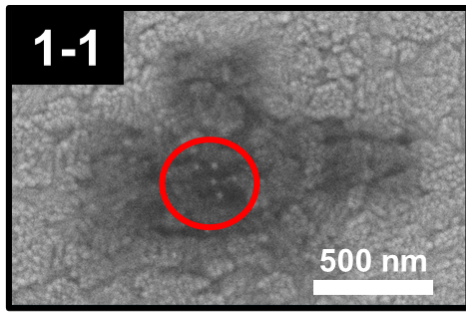
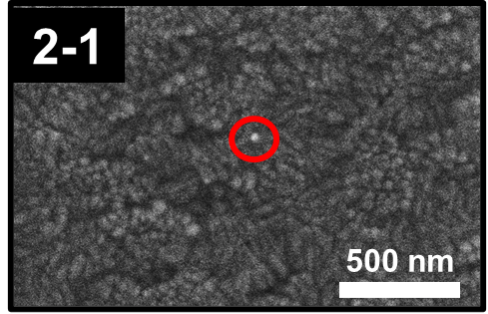
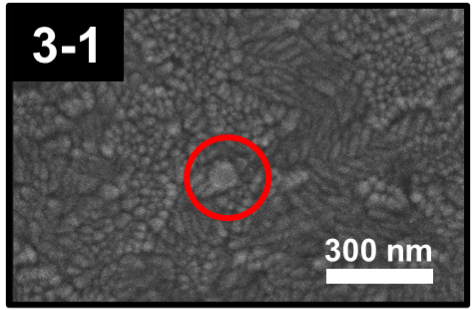
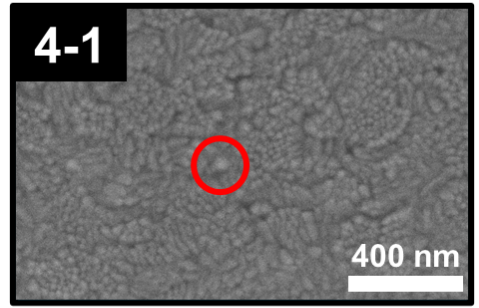
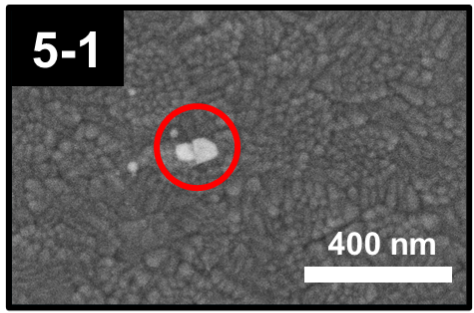
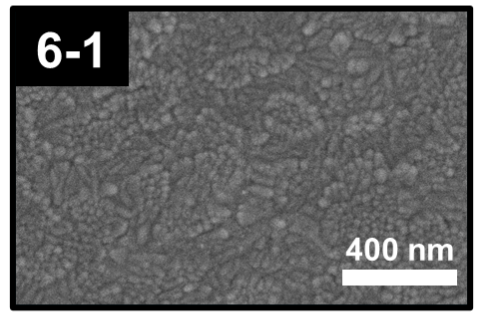
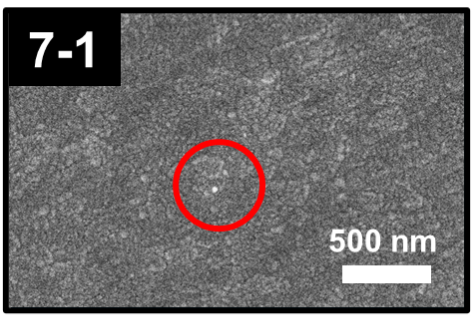
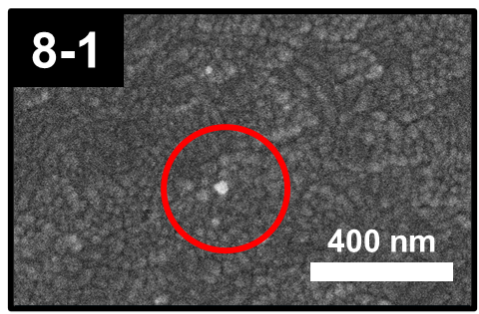
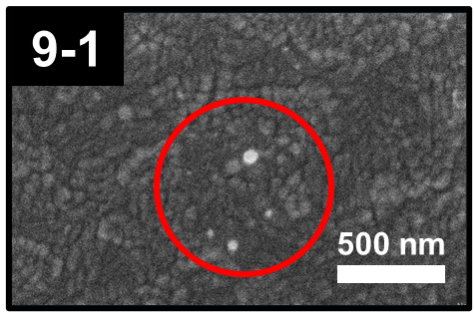
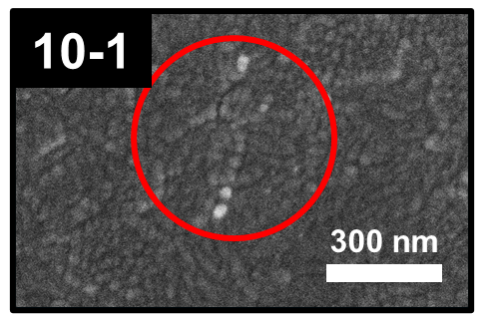

**Extended Data 1 - Row 2**

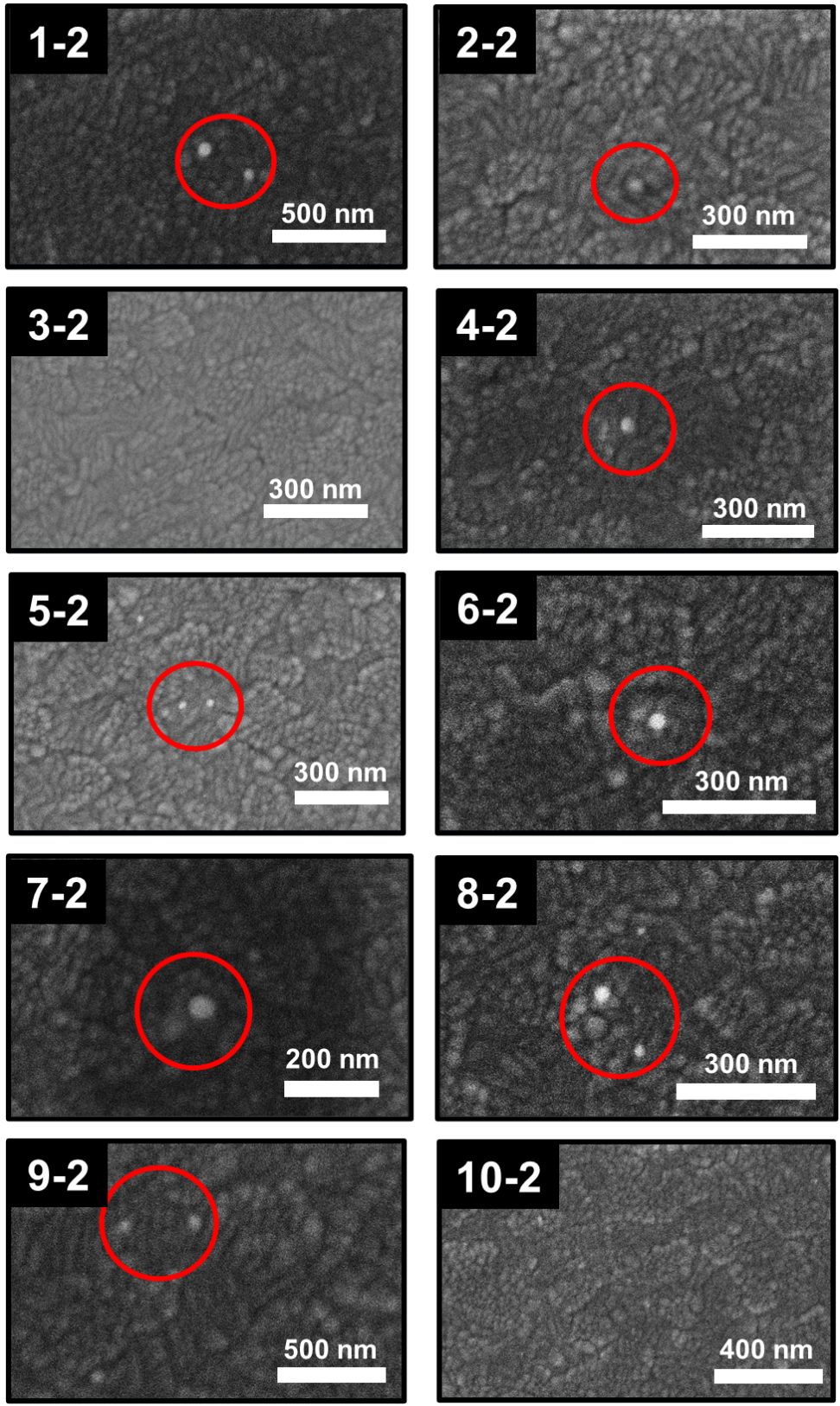

**Extended Data 1 - Row 3**

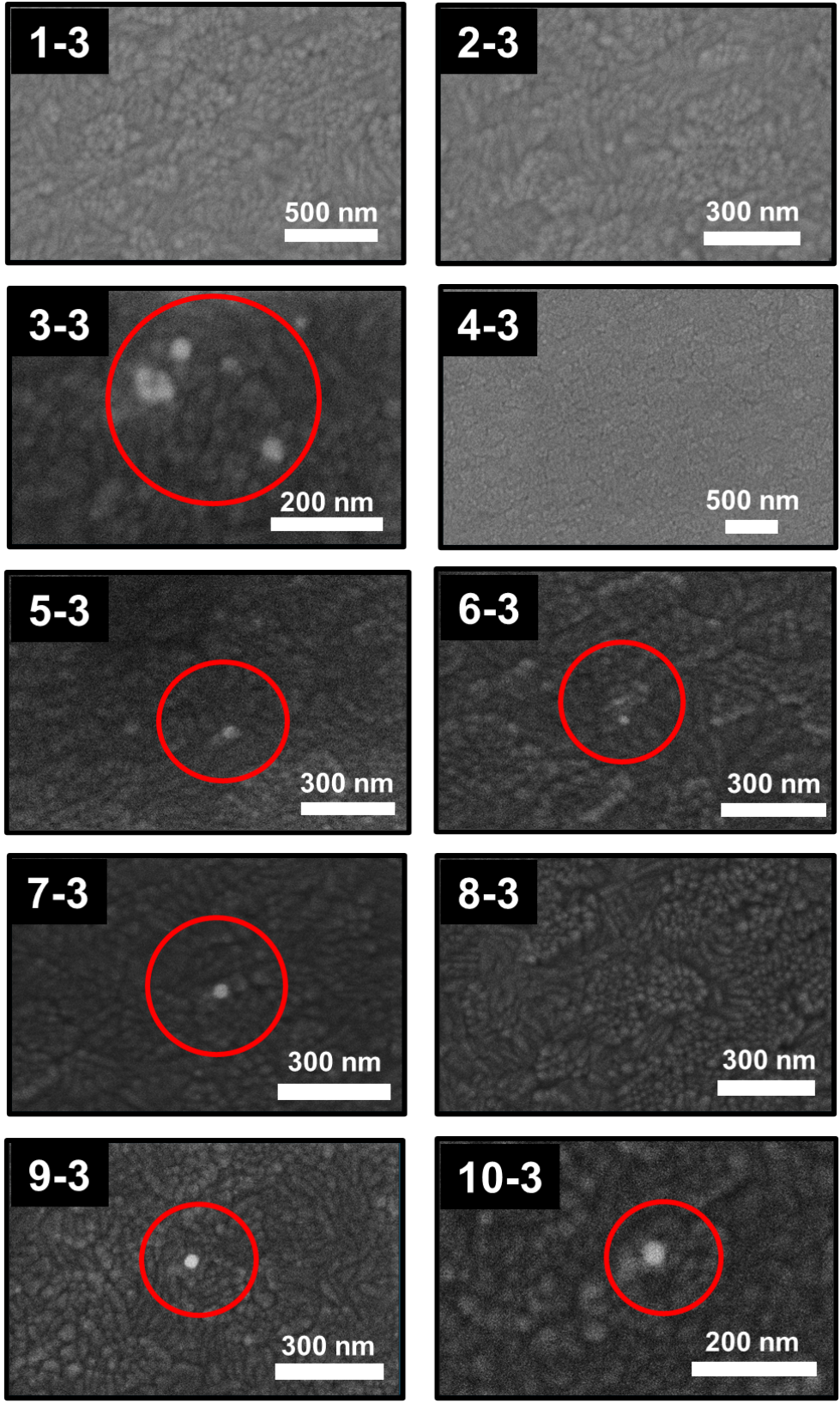

**Extended Data 1 - Row 4**

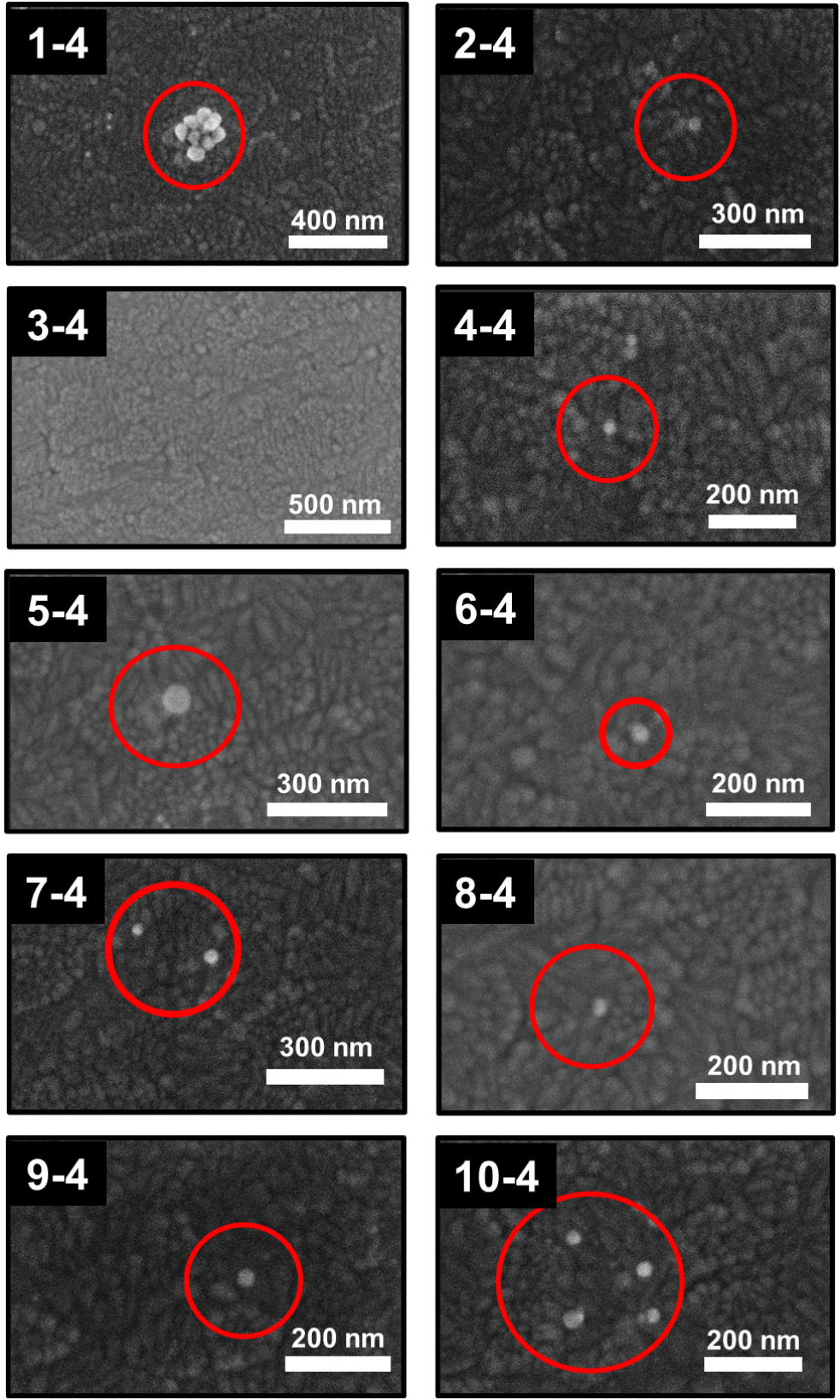

**Extended Data 1 - Row 5**

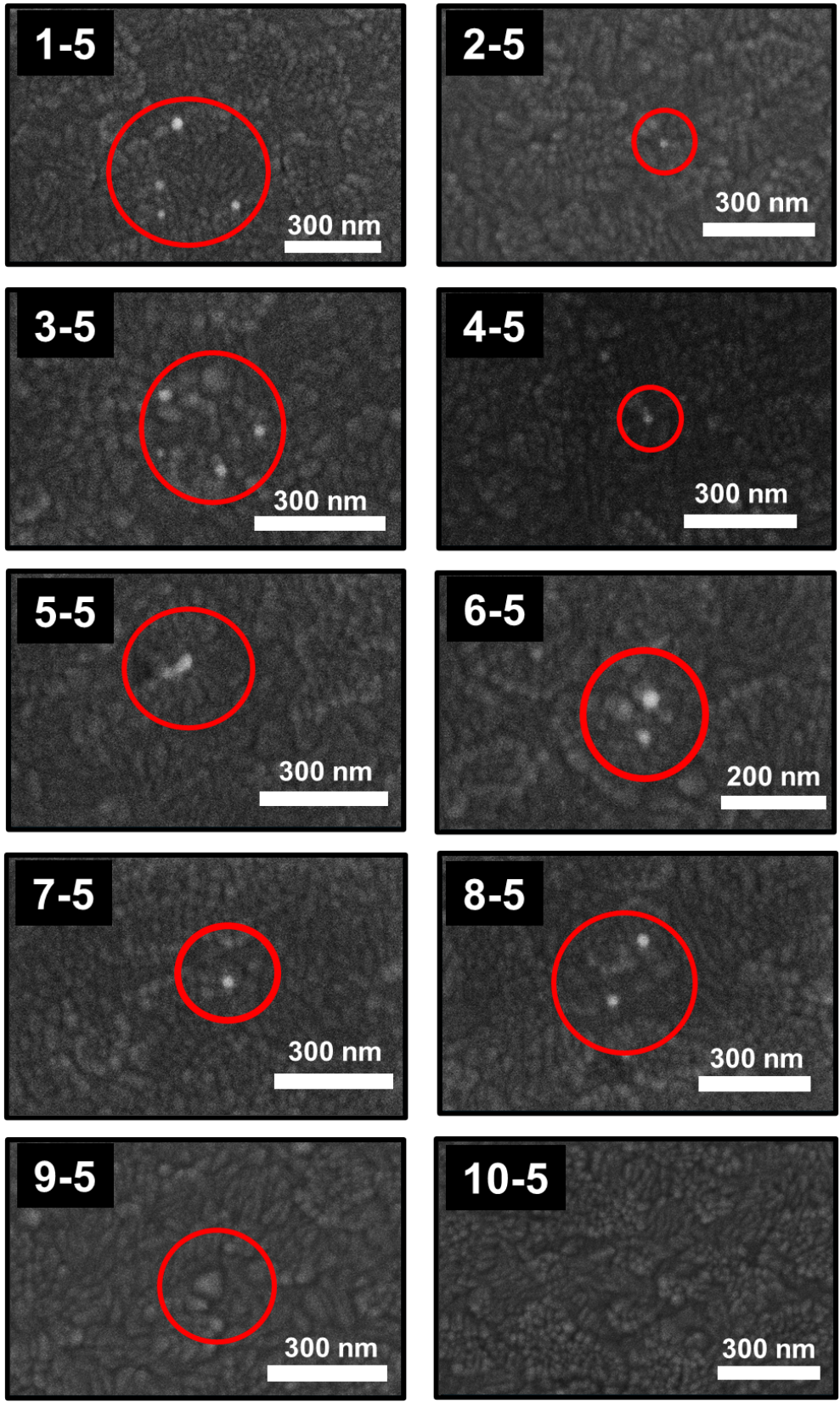

**Extended Data 1 - Row 6**

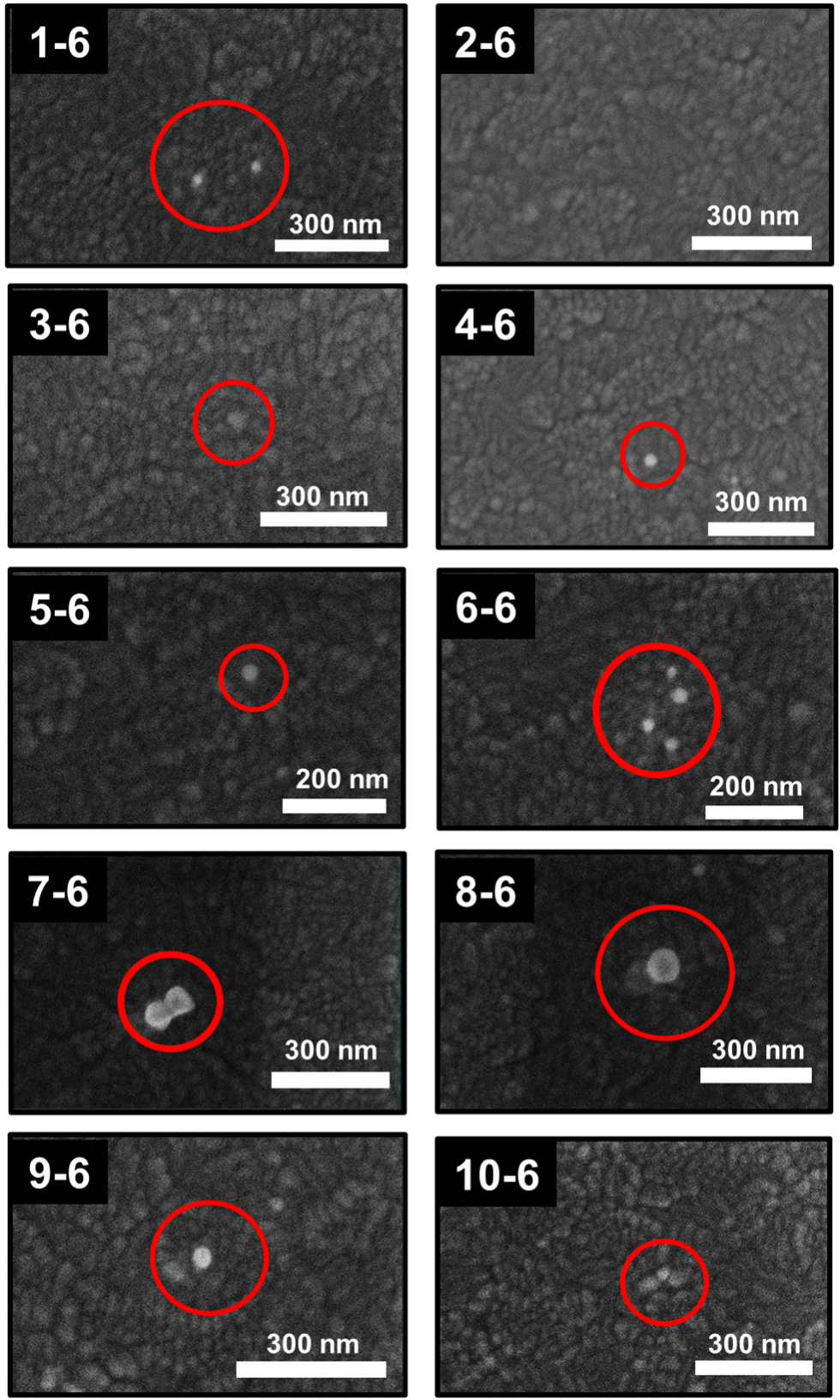

# Extended Data 1 - Row 7

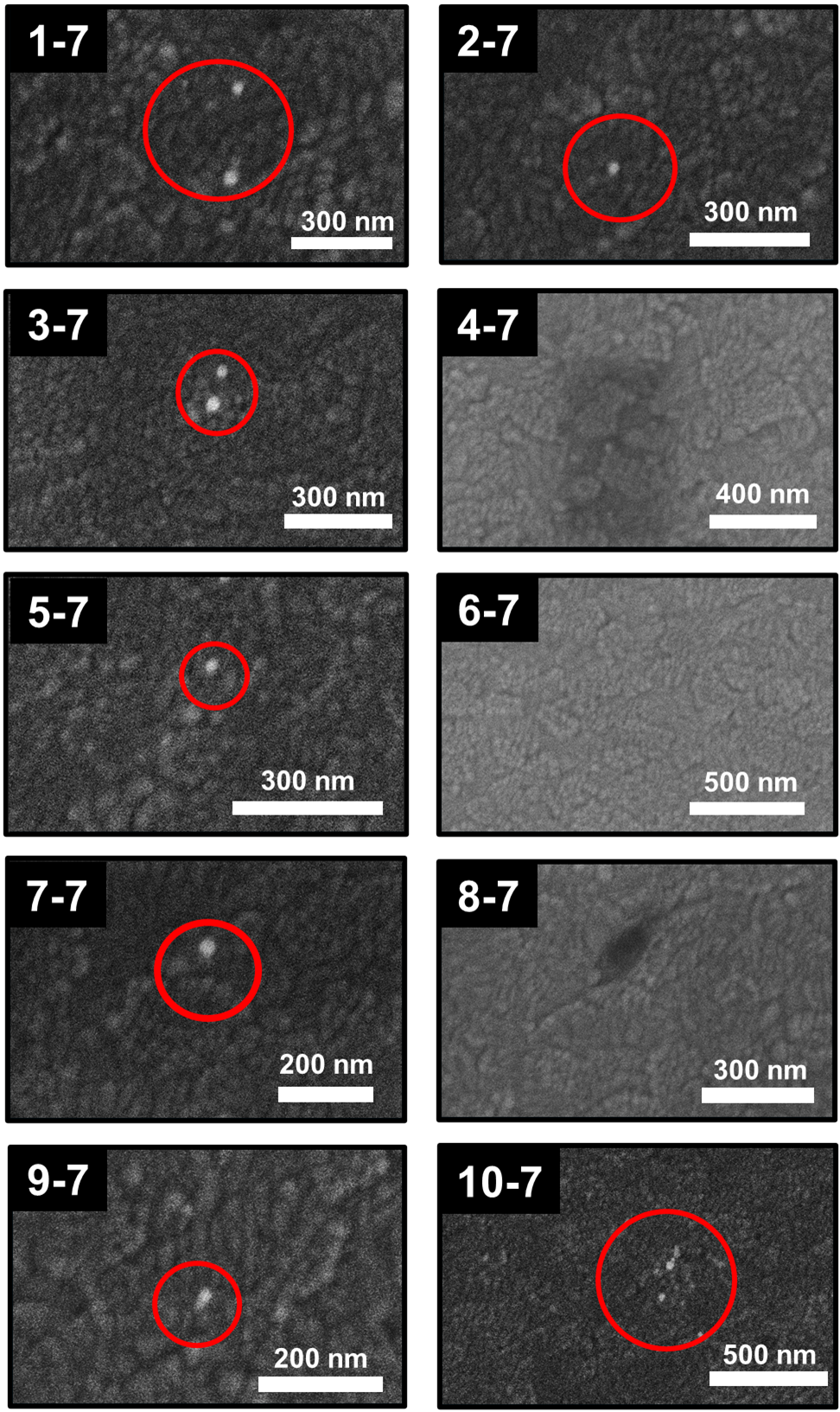

# Extended Data 1 - Row 8

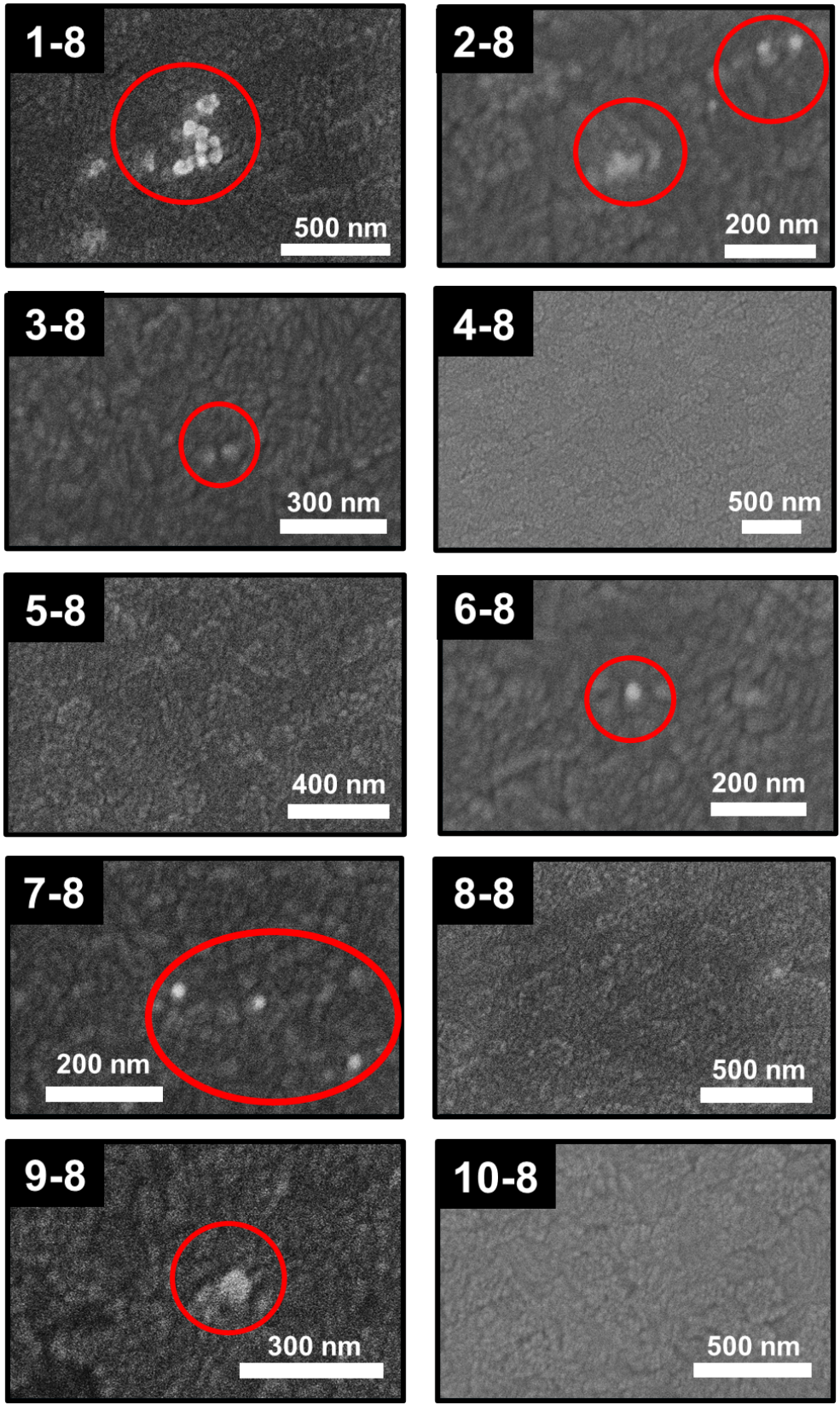

**Extended Data 1 - Row 9**

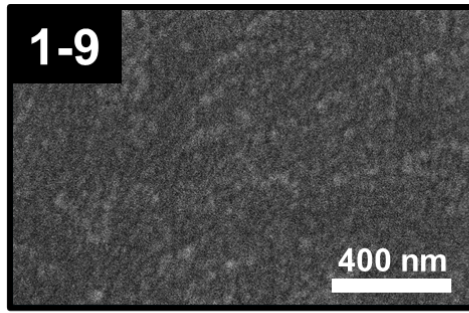
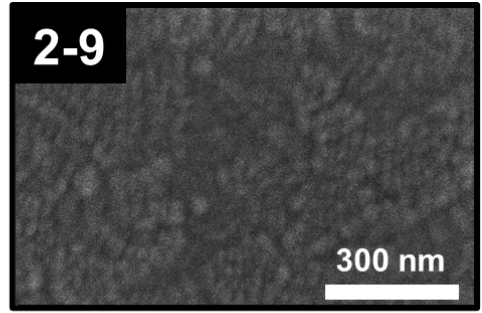
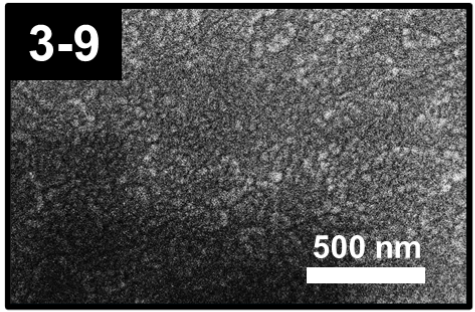
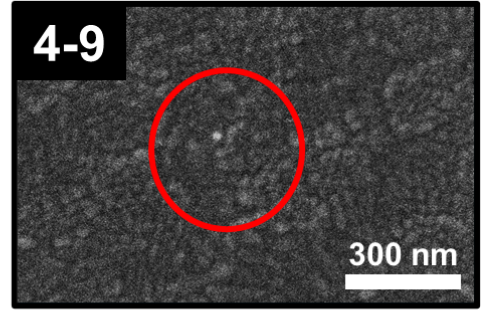
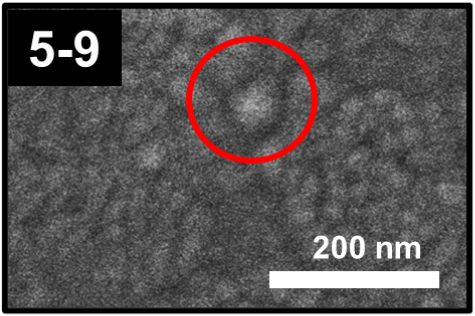
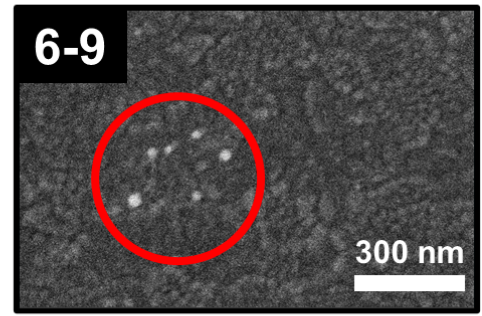
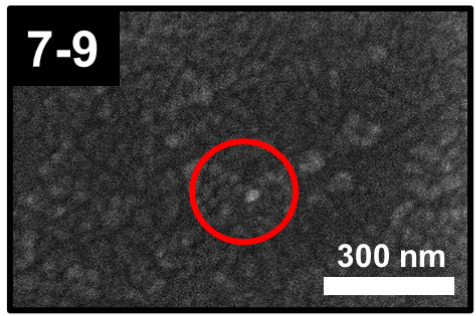
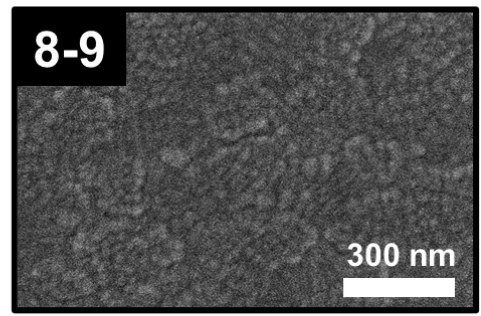
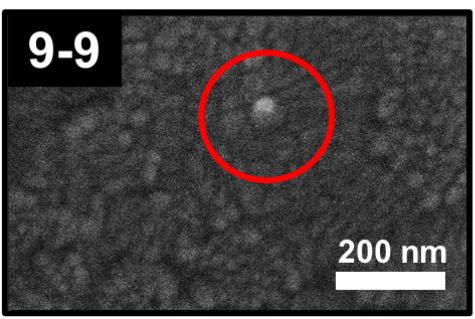
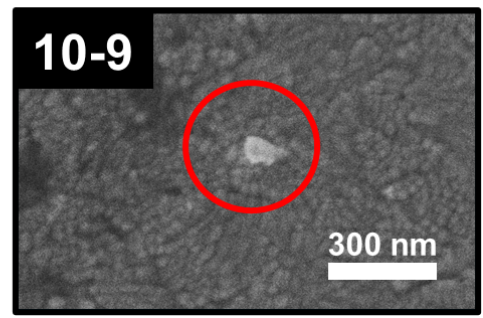

**Extended Data 1 - Row 10**

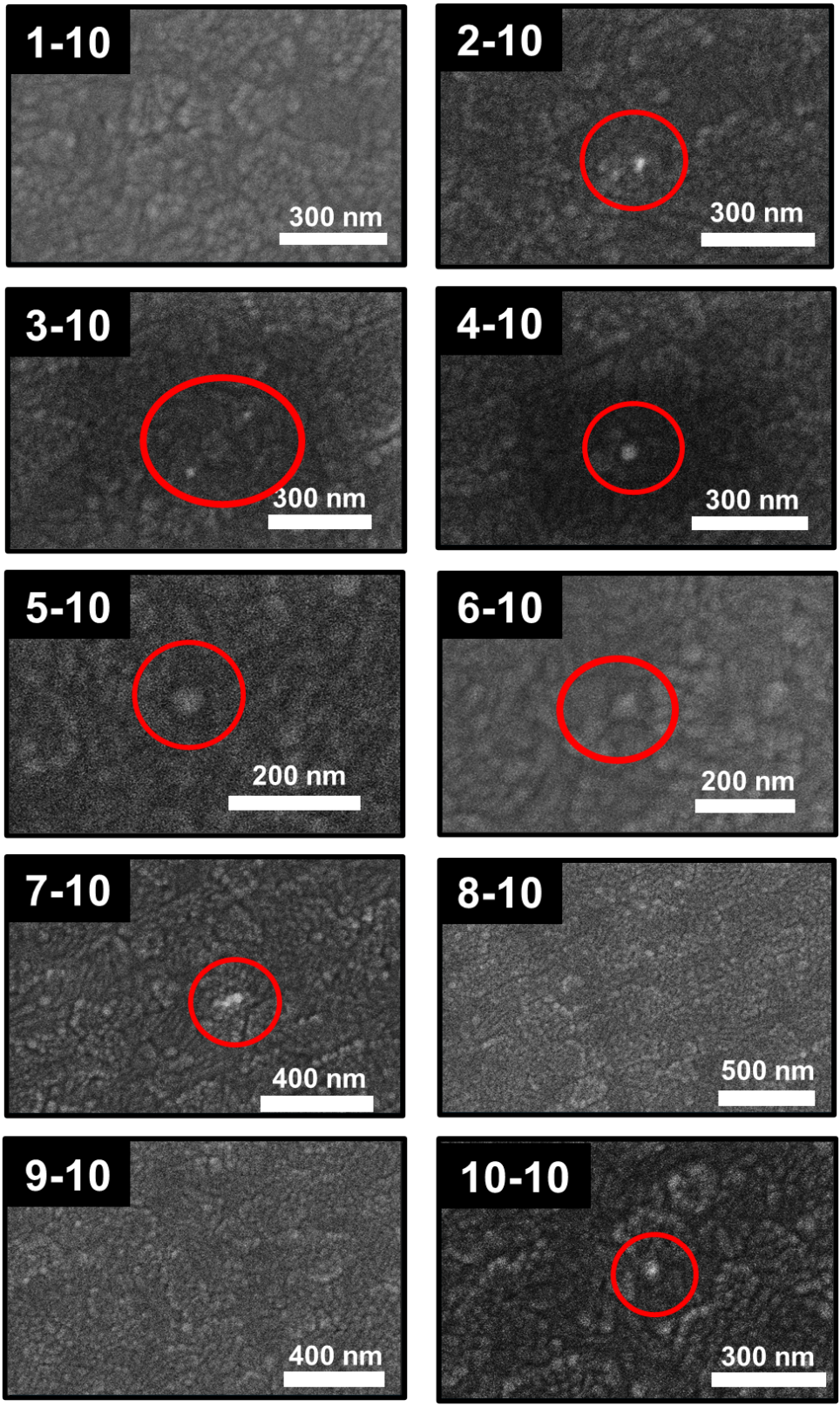